\documentclass[aps,prd,twocolumn,superscriptaddress,nofootinbib]{revtex4-1}

\usepackage{latexsym}
\usepackage{amsmath}
\usepackage{amssymb}
\usepackage{amsfonts}

\usepackage{color}

\usepackage{supertabular} 
\usepackage{placeins}
\usepackage{epsfig}
\usepackage{graphicx}

\def\tstrut{\vrule height3.25ex depth0pt width0pt} 

\begin{document}


\title{Triply-heavy baryons in a constituent quark model}

\author{Gang Yang}
\email[]{ygz0788a@sina.com}
\affiliation{Department of Physics and State Key Laboratory of Low-Dimensional Quantum Physics, \\ Tsinghua University, Beijing 100084, China}

\author{Jialun Ping}
\email[]{jlping@njnu.edu.cn}
\affiliation{Department of Physics and Jiangsu Key Laboratory for Numerical Simulation of Large Scale Complex Systems, \\ Nanjing Normal University, Nanjing 210023, P. R. China}

\author{Pablo G. Ortega}
\email[]{pgortega@usal.es}
\affiliation{Grupo de F\'isica Nuclear and Instituto Universitario de F\'isica Fundamental y Matem\'aticas (IUFFyM), Universidad de Salamanca, E-37008 Salamanca, Spain}

\author{Jorge Segovia}
\email[]{jsegovia@upo.es}
\affiliation{Departamento de Sistemas F\'isicos, Qu\'imicos y Naturales, \\ Universidad Pablo de Olavide, E-41013 Sevilla, Spain}

\date{\today}

\begin{abstract}
A constituent quark model, which has recently been successfully applied to the study of heavy quarkonium properties such as its spectrum but also a diverse array of observables related with their electromagnetic, strong and weak decays and reactions, is used herein to compute ground- and excited-state masses of $QQQ$-baryons containing either $c$- or $b$-quarks.
Considering the lack of experimental information about the spectra of triply-heavy baryons, we believe that our computation could help on finding new states, since it is expected that phenomenological quark models describe triply-heavy baryons to a similar degree of accuracy as heavy quarkonia.
The quark model parameters previously used to describe $c\bar c$ and $b\bar b$ properties have not been modified for this analysis. The non-relativistic three-body bound-state problem is solved by means of the Gau\ss ian expansion method which provides enough accuracy and simplifies the subsequent evaluation of the matrix elements.
Several low-lying states with quantum numbers $J^P=\frac{1}{2}^{\pm}$, $\frac{3}{2}^{\pm}$, $\frac{5}{2}^{\pm}$ and $\frac{7}{2}^+$ are reported. We compare our results with those predicted by many other theoretical formalisms. There is a general trend of agreement about the mass of the ground state in each sector of triply-heavy baryons; however, the situation is more puzzling for the excited states and thus appropriate comments on the most relevant features of our comparison are given.
%
\end{abstract}


\maketitle


\section{Introduction}

Mesons containing only heavy valence quarks, either $c\bar c$ (charmonium) or $b\bar b$ (bottomonium), have contributed to the understanding of Quantum Chromodynamics (QCD) due to their approximately non-relativistic nature and the clean spectrum of narrow states, at least below open-flavor threshold. It is a fact that many precise experimental results are available for conventional heavy quarkonia and their analysis have contributed significantly to the understanding of, for instance, the quark-antiquark forces~\cite{Brambilla:2010cs, Bali:2000gf}. On the other hand, tens of charmonium- and bottomonium-like XYZ states have been identified in experiments at B-factories (BaBar, Belle, and CLEO), $\tau$-charm facilities  (CLEO-c and BES) and hadron colliders (CDF, D0, LHCb, ATLAS, and CMS). So far, there is no conclusion about the nature of these exotic states (see Refs.~\cite{Brambilla:2010cs, Olsen:2014qna, Brambilla:2019esw} for reviews of the experimental and theoretical status of the subject). Their analysis and new determinations will continue with the upgrade of experiments such as BES III~\cite{Asner:2008nq}, Belle II~\cite{Bevan:2014iga} and HL- and HE-LHC~\cite{Cerri:2018ypt}. This will provide a sustained progress in the field as well as the breadth and depth necessary for a vibrant heavy quarkonium research environment.

Being baryonic analogues of heavy quarkonia, triply-heavy baryons may provide a complementary window in the understanding of the strong interaction between quarks without taking into account the usual light-quark complications. Moreover, as in heavy quarkonia, there is no restriction of finding exotic structures in the triply-heavy baryon spectra and thus a reliable prediction of conventional $QQQ$-baryons\footnote{For now on we will denote $Q$ as the heavy quarks $c$ or $b$, being $QQQ$ a baryon composed of any combination of these heavy quarks.} is interesting by itself in order to provide a template from which compare the future experimental findings.

The production of triply-heavy baryons is extremely difficult and thus no experimental signal for any of them has still been reported. Baranov {\it et al.} estimated that triply-charmed baryons may not be observed in $e^+e^-$ collisions and the expectations for $bbb$-baryons would be even worse~\cite{Baranov:2004er}. This conclusion, however, was also reached by Bjorken in the 1980s~\cite{Bjorken:1985a} and he proposed hadron-induced fixed target experiments as the best strategy to follow in order to observe the ground-state triply-charmed baryon, $\Omega_{ccc}^{++}$. First estimates of the cross section production of triply-heavy baryons at LHC were evaluated in Refs.~\cite{Saleev:1999ti, GomshiNobary:2003sf, GomshiNobary:2004mq, GomshiNobary:2005ur}. A more recent calculation~\cite{Chen:2011mb} finds that around $10^4$-$10^5$ events of triply-heavy baryons, with $ccc$ and $ccb$ quark content, can be accumulated for $10\,\text{fb}$ of integrated luminosity. Some investigations on the production rate of multi-charmed hadrons in heavy-ion collisions at high energies have been presented in Refs.~\cite{He:2015hh, Zhao:2017zjx, Cho:2017dcy}. Finally, authors of Refs.~\cite{Flynn:2011gf, Wang:2018ww} have suggested to look for triply-heavy baryons through their semi-leptonic and non-leptonic decays.

From a theoretical point of view, up to our knowledge, the first study of heavy baryon spectroscopy was carried out in Ref.~\cite{Hasenfratz:1980ka} using a QCD-motivated bag model. More recently, there have been other theoretical mass determinations that include non-relativistic quark models~\cite{Bhavin:2009bp, Vijande:2015faa, Zalak:2018zs} and its relativistic variants~\cite{Martynenko:2007je, Migura:2006ep}, Faddeev formalism using a non-relativistic reduction of the quark-quark interaction~\cite{SilvestreBrac:1996bg}, front-form formulation of an effective QCD Hamiltonian~\cite{Serafin:2018aih}, QCD sum rules~\cite{Zhang:2009re, Wang:2011ae, Aliev:2012tt, Aliev:2014lxa}, non-relativistic effective field theories~\cite{Brambilla:2005yk, Jia:2006gw, Brambilla:2009cd, LlanesEstrada:2011kc}, continuum approach to QCD based on Dyson-Schwinger and Faddeev equations~\cite{Qin:2019hgk, Yin:2019bxe}, and lattice gauge theories~\cite{Meinel:2010pw, Meinel:2012qz, Padmanath:2013zfa, Namekawa:2013vu, Zachary:2014zsb, Mathur:2018epb}.

In this work, we shall compute the spectrum of triply-heavy baryons, including ground and excited states with quantum numbers $J^P=\frac{1}{2}^{\pm}$, $\frac{3}{2}^{\pm}$, $\frac{5}{2}^{\pm}$ and $\frac{7}{2}^+$. Our theoretical formalism is the constituent quark model (CQM) proposed in Ref.~\cite{Vijande:2004he} (see references~\cite{Valcarce:2005em} and~\cite{Segovia:2013wma} for reviews). This model has recently been successfully applied to mesons containing heavy quarks, studying their spectra~\cite{Segovia:2008zz, Segovia:2010zzb, Segovia:2011tb, Segovia:2016xqb}, their electromagnetic, weak and strong decays and reactions~\cite{Segovia:2008zz, Segovia:2016xqb, Segovia:2011dg, Segovia:2011zza, Segovia:2012cd, Segovia:2013kg, Segovia:2014mca}, their coupling with meson-meson thresholds~\cite{Ortega:2009hj, Ortega:2016hde, Ortega:2017qmg, Ortega:2018cnm} and, lately, the phenomenological exploration of multiquark structures~\cite{Yang:2015bmv, Yang:2018oqd}. Therefore, this study entails the first step towards an unified description of heavy mesons and baryons using the same quark model, parameters and technical formalism presented in Refs.~\cite{Segovia:2008zz, Segovia:2016xqb}. Moreover, the predicted spectrum of triply-heavy baryons could also be seen as another contribution to the template against which to compare the future experimental findings, and discern between conventional and exotic structures because potential models are expected to describe triply-heavy baryons to a similar degree of accuracy as the successful results obtained in the charmonium and bottomonium sectors.

The present manuscript is arranged as follows. We describe briefly in Sec.~\ref{sec:model} the constituent quark model, the triply-heavy baryon wave-function and the computational formalism based on the Gau\ss ian expansion method. Section~\ref{sec:results} is devoted to the analysis and discussion of the obtained results. We summarize and give some prospects in Sec.~\ref{sec:summary}.


\section{Theoretical framework}
\label{sec:model}

The Hamiltonian which describes the triply-heavy baryon bound-state system can be written as
\begin{equation}
H = \sum_{i=1}^{3}\left( m_i+\frac{\vec{p\,}^2_i}{2m_i}\right) - T_{\text{CM}} + \sum_{j>i=1}^{3} V(\vec{r}_{ij}) \,.
\label{eq:Hamiltonian}
\end{equation}
where $T_{\text{CM}}$ is the center-of-mass kinetic energy. Since chiral symmetry is explicitly broken in the heavy quark sector, the two-body potential can be deduced from the one-gluon exchange and confining interactions. The one-gluon exchange potential is given by
\begin{align}
&
V_{\text{OGE}}(\vec{r}_{ij}) = \frac{1}{4} \alpha_{s} (\vec{\lambda}_{i}^{c}\cdot
\vec{\lambda}_{j}^{c}) \Bigg[\frac{1}{r_{ij}} \nonumber \\ 
&
\hspace*{1.60cm} - \frac{1}{6m_{i}m_{j}} (\vec{\sigma}_{i}\cdot\vec{\sigma}_{j}) 
\frac{e^{-r_{ij}/r_{0}(\mu)}}{r_{ij}r_{0}^{2}(\mu)} \Bigg] \,,
\end{align}
where $m_{i}$ is the quark mass, $\lambda^c$ are the $SU(3)$-color Gell-Mann matrices, and the Pauli matrices are denoted by $\vec{\sigma}$. The contact term of the central potential has been regularized as
\begin{equation}
\delta(\vec{r}_{ij})\sim\frac{1}{4\pi r_{0}^{2}}\frac{e^{-r_{ij}/r_{0}}}{r_{ij}} \,,
\end{equation}
with $r_{0}(\mu_{ij})=\hat{r}_{0} \mu_{nn}/\mu_{ij}$ a regulator that depends on the reduced mass of the quark-quark pair, $\mu_{ij}$; being $\mu_{nn}=(313/2)\,\text{MeV}$ the one corresponding to the lightest quark-quark couple.

The wide energy range needed to provide a consistent description of light, strange and heavy mesons requires an effective scale-dependent strong coupling constant. We use the frozen coupling constant~\cite{Vijande:2004he}
\begin{equation}
\alpha_{s}(\mu_{ij})=\frac{\alpha_{0}}{\ln\left(
\frac{\mu_{ij}^{2}+\mu_{0}^{2}}{\Lambda_{0 }^{2}} \right)},
\end{equation}
in which $\alpha_{0}$, $\mu_{0}$ and $\Lambda_{0}$ are parameters of the model.

Color confinement should be encoded in the non-Abelian character of QCD. Studies on a lattice have demonstrated that multi-gluon exchanges produce an attractive linearly rising potential proportional to the distance between infinitely heavy quarks~\cite{Bali:2005fu}. However, the spontaneous creation of light-quark pairs from the QCD vacuum may give rise at the same scale to a breakup of the created color flux-tube~\cite{Bali:2005fu}. We have tried to mimic these two phenomenological observations by the expression:
\begin{equation}
V_{\text{CON}}(\vec{r}_{ij}\,)=\left[-a_{c}(1-e^{-\mu_{c}r_{ij}})+\Delta \right] 
(\vec{\lambda}_{i}^{c}\cdot\vec{\lambda}_{j}^{c}) \,,
\label{eq:conf}
\end{equation}
where $a_{c}$ and $\mu_{c}$ are model parameters. One can see in Eq.~\eqref{eq:conf} that the potential is linear at short inter-quark distances with an effective confinement strength $\sigma = -a_{c} \, \mu_{c} \, (\vec{\lambda}^{c}_{i}\cdot \vec{\lambda}^{c}_{j})$, while it becomes constant at large distances. 

Let us mention here that associated tensor and spin-orbit terms of the potentials presented above appear not to be essential for a global description of baryons~\cite{SilvestreBrac:1996bg}. Therefore, they have been neglected since our main purpose herein is to get, within our approach, a first and reliable unified description of heavy mesons and baryons.

The quark model parameters relevant for this work are shown in Table~\ref{tab:parameters}. As mentioned above, they were fitted attending to heavy meson phenomenology; see, for instance, Refs.~\cite{Segovia:2008zz, Segovia:2016xqb}.

\begin{table}[!t]
\caption{\label{tab:parameters} Quark model parameters.}
\begin{ruledtabular}
\begin{tabular}{cccc}
Quark masses & $m_c$ (MeV) & $1763$ \\
             & $m_b$ (MeV) & $5110$ \\[2ex]
OGE          & $\hat{r}_{0}$ (fm)               & $0.181$ \\
             & $\alpha_{0}$                     & $2.118$ \\
             & $\Lambda_{0}$ $(\mbox{fm}^{-1})$ & $0.113$ \\
             & $\mu_{0}$ (MeV)                  & $36.976$ \\[2ex]
Confinement  & $a_{c}$ (MeV)                & $507.4$ \\
             & $\mu_{c}$ $(\mbox{fm}^{-1})$ & $0.576$ \\
             & $\Delta$ (MeV)               & $184.432$ \\
\end{tabular}
\end{ruledtabular}
\end{table}

The triply-heavy baryon wave function is constructed from a product of four terms: color, flavor, spin and space wave functions. Concerning the color one, it can be written as usually for a baryon:
\begin{equation}
\chi^{c} = \frac{1}{\sqrt{6}} (rgb - rbg + gbr - grb + brg - bgr) \,.
\end{equation}
The spin wave-function of a 3-quark system has been worked out in, e.g., Ref.~\cite{Yang:2017qan} and, for clarity purposes, we repeat our expressions herein:
\begin{align}
\chi_{\frac32,\frac32}^{\sigma}(3)  &= \alpha\alpha\alpha \,, \\
\chi_{\frac32,\frac12}^{\sigma}(3)  &= \frac{1}{\sqrt{3}}(\alpha\alpha\beta+\alpha\beta\alpha+\beta\alpha\alpha) \,, \\
\chi_{\frac32,-\frac12}^{\sigma}(3) &= \frac{1}{\sqrt{3}} (\alpha\beta\beta+\beta\alpha\beta+\beta\beta\alpha) \,, \\
\chi_{\frac32,-\frac32}^{\sigma}(3)  &= \beta\beta\beta \,, \\
\chi_{\frac12,\frac12}^{\sigma 1}(3) &= \frac{1}{\sqrt{6}} (2\alpha\alpha\beta-\alpha\beta\alpha-\beta\alpha\alpha) \,, \\
\chi_{\frac12,\frac12}^{\sigma 2}(3) &= \frac{1}{\sqrt{2}} (\alpha\beta\alpha-\beta\alpha\alpha) \,, \\
\chi_{\frac12,-\frac12}^{\sigma 1}(3) &= \frac{1}{\sqrt{6}}  (\alpha\beta\beta+\beta\alpha\beta-2\beta\beta\alpha) \,, \\
\chi_{\frac12,-\frac12}^{\sigma 2}(3) &= \frac{1}{\sqrt{2}} (\alpha\beta\beta-\beta\alpha\beta) \,,
\end{align}
being $\alpha=\left|\tfrac{1}{2}\tfrac{1}{2}\right\rangle$ and $\beta=\left|\tfrac{1}{2}-\tfrac{1}{2}\right\rangle$ the spin states of the constituent quarks inside the baryon. The flavor wave-function of a fully heavy quark baryon is trivial
\begin{align}
\chi^f_{ccc} = ccc \,, \\
\chi^f_{ccb} = ccb \,, \\ 
\chi^f_{cbb} = cbb \,, \\
\chi^f_{bbb} = bbb \,.
\end{align}

The spatial wave function of the $3$-body system can be written as a sum of amplitudes of three rearrangement channels
\begin{equation}
\psi_{LM_L} = 
\Phi_{LM_L}^{(c=1)}(\vec{\rho}_1,\vec{\lambda}_1) + \Phi_{LM_L}^{(c=2)}(\vec{\rho}_2,\vec{\lambda}_2) + \Phi_{LM_L}^{(c=3)}(\vec{\rho}_3,\vec{\lambda}_3)
\label{eq:SWF}
\end{equation}
where $\vec{\rho}_i$ and $\vec{\lambda}_i$ are the internal Jacobi coordinates
\begin{equation}
\vec{\rho}_i = \vec{x}_j - \vec{x}_k \,, \quad
\vec{\lambda}_i = \vec{x}_i - \frac{m_j \vec{x}_j + m_k \vec{x}_k}{m_j+m_k} \,,
\end{equation}
with $i,\,j,\,k=1,\ldots,3$ and $i\neq j\neq k$. Note here that we shall work with a triply-heavy baryon in which either the three quarks are the same and then only one rearrangement channel is needed; or two of the three quarks are equal and thus two rearrangement channels must been incorporated.

Each amplitude in Eq.~\eqref{eq:SWF} is expanded in terms of infinitesimally-shifted Gaussian basis functions~\cite{Hiyama:2003cu}:
\begin{equation}
\Phi_{LM_L}^{(c)}(\vec{\rho}_c,\vec{\lambda}_c) = \sum_{n_1l_1,n_2l_2} A_{n_1l_1,n_2l_2}^{(c)} \Big[ \phi_{n_1l_1}(\vec{\rho}_c) \, \varphi_{n_2l_2}(\vec{\lambda}_c) \Big]_{LM_L}
\label{eq:l1l2L}
\end{equation}
where
\begin{align}
&
\phi_{n_1l_1m_1}(\vec{\rho}\,) = N_{n_1l_1} \rho^{l_1} e^{-\nu_{n_1} \rho^2} Y_{l_1m_1}(\hat{\rho}) \nonumber \\
&
\hspace*{0.60cm} = N_{n_1l_1} \lim_{\varepsilon\to 0} \frac{1}{(\nu_{n_1}\varepsilon)^{l_1}} \sum_{k=1}^{k_{\rm
max}} C_{l_1m_1,k} \, e^{-\nu_{n_1}(\vec{\rho}-\varepsilon \vec{D}_{l_1m_1,k})^{2}} \,, \\
&
\varphi_{n_2l_2m_2}(\vec{\lambda}\,) = N_{n_2l_2} \lambda^{l_2} e^{-\nu_{n_2} \lambda^2} Y_{l_2m_2}(\hat{\lambda}) \nonumber \\
&
\hspace*{0.60cm} = N_{n_2l_2} \lim_{\varepsilon\to 0} \frac{1}{(\nu_{n_2}\varepsilon)^{l_2}} \sum_{t=1}^{t_{\rm max}} C_{l_2m_2,t} \, e^{-\nu_{n_2}(\vec{\lambda}-\varepsilon \vec{D}_{l_2m_2,t})^{2}} \,.
\end{align}
The spherical harmonics are denoted by $Y_{l_1m_1}(\hat{\rho})$ and $Y_{l_2m_2}(\hat{\lambda})$; $N_{n_1l_1}$ and $N_{n_2l_2}$ are normalization constants; and the determination of the basis parameters \{$C_{l_1m_1,k}$, $D_{l_1m_1,k}$; $k=1,\ldots,k_{\max}$\} as well as \{$C_{l_2m_2,t}$, $D_{l_2m_2,t}$; $t=1,\ldots,t_{\max}$\} is described in, for instance, Appendix A.2 of Ref.~\cite{Hiyama:2003cu}. The limit $\varepsilon\to 0$ must be carried out after the matrix elements have been calculated analytically. This new set of basis functions makes the calculation of $3$-body matrix elements easier without the laborious Racah algebra. Following Ref.~\cite{Hiyama:2003cu}, the Gau\ss ian ranges $\nu_{n_i}$ with $i=1,\,2$ are taken in a geometric progression. This enables their optimization employing a small number of free parameters. Moreover, the geometric progression is dense at short distances, so that it allows the description of the dynamics mediated by short range potentials. The fast damping of the Gau\ss ian tail is not a problem, since we can choose the maximal range much longer than the hadronic size.

The Rayleigh-Ritz variational principle is used in order to solve the Schr\"odinger equation
\begin{equation}
\big[ H-E \big] \, \Psi_{JM_J} = 0 \,,
\end{equation}
and determine the eigenenergies, $E$, and coefficients $A_{n_1l_1,n_2l_2}^{(c)}$. Note herein that the complete wave-function is written as
\begin{equation}
\Psi_{JM_{J}}={\cal A} \left\{ \left[ \psi_{LM_L} \chi^{\sigma}_{SM_S}(3) \right]_{JM_J} \chi^{f} \chi^{c} \right\} \,,
\label{wf}
\end{equation}
where, in order to fulfill the Pauli principle, the antisymmetric operator $\cal A$ is the same for $\Omega_{ccc}$ and $\Omega_{bbb}$, i.e. ${\cal A}=1-(13)-(23)$ in a system with three identical particles.\footnote{Note here that each (ij)-term in ${\cal A}$ express an interchange operator of $S_3$ permutation group for $QQQ$-clusters, with $Q$ either a $c$- or $b$-quark.} However, ${\cal A}=1$ for $\Omega_{ccb}$ and ${\cal A}=1-(23)$ for $\Omega_{cbb}$. This is needed because we have constructed an antisymmetric wave function for only the first two quarks of the 3-quark cluster, the remaining quark of the system has been added to the wave function by simply considering the appropriate Clebsch-Gordan coefficient.


\begin{table}[!t]
\caption{\label{tab:couplings} Possible $J^P$ quantum numbers of all studied $3$-heavy-quark bound-state systems.}
\begin{ruledtabular}
\begin{tabular}{cccc}
$(l_1,l_2)$& $L$ & $S=\frac{1}{2}$ & $S=\frac{3}{2}$ \\[2ex]
$(0,0)$& $0$ & $\frac{1}{2}^+$ & $\frac{3}{2}^+$ \\[1ex]
$\begin{matrix} (0,1) \\ (1,0) \end{matrix}$ & $1$ & $\frac{1}{2}^-$, $\frac{3}{2}^-$ & $\frac{1}{2}^-$, $\frac{3}{2}^-$, $\frac{5}{2}^-$ \\[3ex]
$\begin{matrix} (0,2) \\ (1,1) \\ (2,0) \end{matrix}$ & $2$ & $\frac{1}{2}^+$, $\frac{3}{2}^+$, $\frac{5}{2}^+$ & $\frac{1}{2}^+$, $\frac{3}{2}^+$, $\frac{5}{2}^+$, $\frac{7}{2}^+$ \\
\end{tabular}
\end{ruledtabular}
\end{table}

\section{Results}
\label{sec:results}

Table~\ref{tab:couplings} shows the total spin and parity, $J^P$, of the triply-heavy baryons whose masses shall be calculated. In a non-relativistic approximation, the total angular momentum, $L$, and spin, $S$, are good quantum numbers and they couple to total spin $J$.\footnote{Since spin-orbital interactions are not employed in the present work, triply-heavy baryons with different $J^P$ but equal $L$ and $S$ will be degenerated in mass.} The total angular momentum is reached by the coupling of the two possible excitations along the Jacobi coordinates, i.e. the $l_1$ and $l_2$ of Eq.~\eqref{eq:l1l2L} couple to $L$. In this analysis, the $l$'s shall never be greater than $L$ and the possible channels have been listed in the first column of Table~\ref{tab:couplings}. Since a baryon is a $3$-quark bound-state system, its total spin can only take values $1/2$ and $3/2$ and its parity is given by $(-1)^{l_1+l_2}$, since the parity of a quark is positive by convention.

Tables~\ref{tab:Eccc},~\ref{tab:Eccb},~\ref{tab:Ecbb} and~\ref{tab:Ebbb} show, respectively, the spectrum of $\Omega_{ccc}$, $\Omega_{ccb}$, $\Omega_{cbb}$ and $\Omega_{bbb}$ baryon sectors computed within our formalism and following the pattern of quantum numbers discussed above and shown in Table~\ref{tab:couplings}. Since there is no experimental data related with triply-heavy baryons, we compare first our results with those predicted by lattice QCD when available. However, lattice-regularized computations have their own issues such as the use of non-relativistic QCD (NRQCD) actions for heavy quarks which are not ideally suited for charm quarks and the fact of not addressing all the systematic uncertainties. For this reason, we shall add for each sector a table that shows mass predictions of other theoretical approaches and compare them with our results (Tables~\ref{tab:Cccc},~\ref{tab:Cccb},~\ref{tab:Ccbb}, and~\ref{tab:Cbbb}).


\begin{table}[!t]
\caption{\label{tab:Eccc} Predicted masses, in MeV, of $\Omega_{ccc}$ baryons with total spin and parity $J^P=\frac{1}{2}^\pm$, $\frac{3}{2}^{\pm}$, $\frac{5}{2}^\pm$ and $\frac{7}{2}^+$. We compare our results with those obtained by lattice QCD in Ref.~\cite{Padmanath:2013zfa}; note that only statistical uncertainties are given by lattice.}
\begin{ruledtabular}
\begin{tabular}{cccc}
$J^P$ & $nL$ & This work & Ref.~\cite{Padmanath:2013zfa} \\
\hline
\tstrut
$\frac{1}{2}^+$ & $1D$ & $5376$ & $5395\pm13$ \\
& $2D$ & $5713$ & - \\
$\frac{3}{2}^+$ & $1S$ & $4798$ & $4759\pm6$  \\
& $2S$ & $5286$ & $5313\pm31$ \\
$\frac{3}{2}^+$ & $1D$ & $5376$ & $5426\pm13$ \\
& $2D$ & $5713$ & - \\
$\frac{5}{2}^+$ & $1D$ & $5376$ & $5402\pm15$ \\
& $2D$ & $5713$ & - \\
$\frac{7}{2}^+$ & $1D$ & $5376$ & $5393\pm49$ \\
& $2D$ & $5713$ & -             \\[0.60ex]
\hline
\tstrut
$\frac{1}{2}^-$ & $1P$ & $5129$ & $5116\pm9$  \\
& $2P$ & $5525$ & $5608\pm31$ \\
$\frac{3}{2}^-$ & $1P$ & $5129$ & $5120\pm13$ \\
& $2P$ & $5525$ & $5658\pm31$ \\
$\frac{5}{2}^-$ & $1P$ & $5558$ & $5512\pm64$ \\
& $2P$ & $5846$ & $5705\pm25$ \\
\end{tabular}
\end{ruledtabular}
\end{table}

\begin{figure*}[ht]
\centering
\includegraphics[clip,trim={0.0cm 0.0cm 0.0cm 0.0cm},width=0.8\textwidth, height=0.43\textheight]{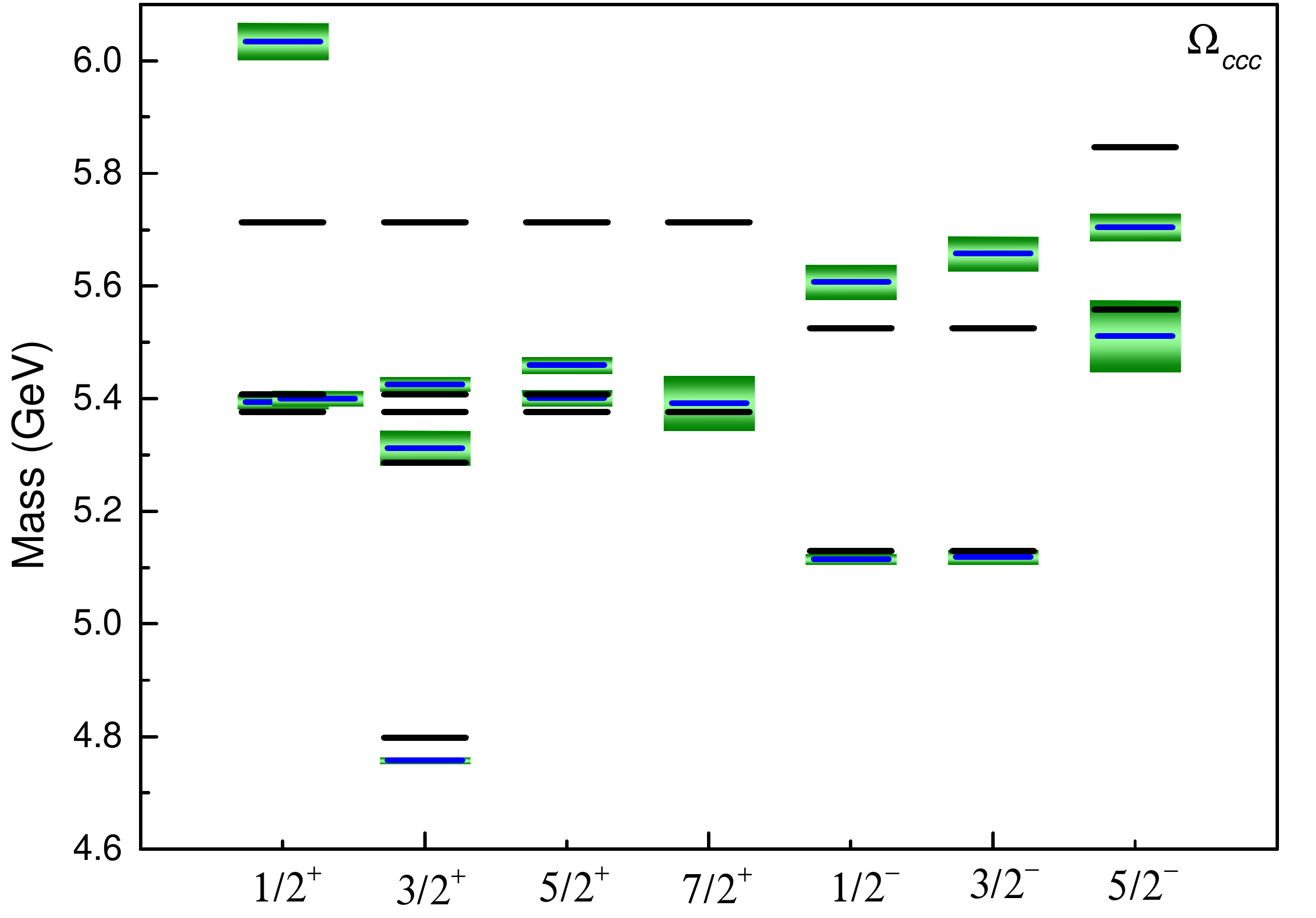}
\caption{\label{fig:Eccc} Spectrum of $\Omega_{ccc}$ baryons with total spin and parity $J^P=\frac{1}{2}^\pm$, $\frac{3}{2}^{\pm}$, $\frac{5}{2}^\pm$ and $\frac{7}{2}^+$. We compare our results (black solid lines) with those obtained by lattice QCD in Ref.~\cite{Padmanath:2013zfa} (blue lines with green boxes); note that only statistical uncertainties are given by lattice.}
\end{figure*}

\begin{table*}[!t]
\caption{\label{tab:Cccc} Predicted masses, in MeV, of $\Omega_{ccc}$ baryons by other theoretical approaches compared with our results.} 
\begin{ruledtabular}
\begin{tabular}{ccccccc}
$nL\, (J^P)$ & $1S\, (\frac{3}{2}^+)$ & $2S\, (\frac{3}{2}^+)$ & $1D\, (\frac{1}{2}^+)$ & $1P\, (\frac{1}{2}^-)$ & $1P\, (\frac{3}{2}^-)$ & $1P\, (\frac{5}{2}^-)$ \\[0.60ex]
\hline
This work                   & $4798$ & $5286$ & $5376$ & $5129$ & $5129$ &$5558$ \\
\cite{Bhavin:2009bp}        & $4812\pm85$ & - & - & - & - & - \\
\cite{Vijande:2015faa}      & $4763$ & $5317$ & $5412$ & $5132$ & $5132$ & $5637$ \\
\cite{Martynenko:2007je}    & $4803$ & - & - & - & - & - \\ 
\cite{Migura:2006ep}        & $4773$ & - & $5216$ & $5109$ & $5014$ & - \\
\cite{SilvestreBrac:1996bg} & $4799$ & $5243$ & $5324$ & $5094$ & $5094$ & $5494$ \\
\cite{Serafin:2018aih}      & $4797$ & $5309$ & $5358$ & $5103$ & $5103$ & - \\
\cite{Zhang:2009re}         & $4670\pm150$ & - & - & - & - & - \\
\cite{Wang:2011ae}          & $4990\pm140$ & - & - & - & $5110\pm150$ & - \\
\cite{Aliev:2014lxa}        & $4720\pm120$ & - & - & - & $4900\pm100$ & - \\ 
\cite{Jia:2006gw}           & $4760\pm60$ & - & - & - & - & - \\
\cite{LlanesEstrada:2011kc} & $4979\pm271$ & - & - & - & - & - \\
\cite{Qin:2019hgk}          & $4760$ & $5150$ & - & - & $5027$ & - \\
\cite{Yin:2019bxe}          & $5000$ & - & - & - & - & - \\
\cite{Namekawa:2013vu}      & $4789\pm6\pm21$ & - & - & - & - & - \\
\cite{Zachary:2014zsb}      & $4796\pm8\pm18$ & - & - & - & - & - \\
\cite{Vijande:2004at}       & $4632$ & - & $4915\pm283$ & $4808\pm176$ & - & - \\
\end{tabular}
\end{ruledtabular}
\end{table*}

{\bf The $\mathbf{\Omega_{ccc}}$ baryon sector.} Table~\ref{tab:Eccc} and Fig.~\ref{fig:Eccc} compare our predicted masses of $\Omega_{ccc}$ baryons with total spin and parity $J^P=\frac{1}{2}^\pm$, $\frac{3}{2}^{\pm}$, $\frac{5}{2}^\pm$ and $\frac{7}{2}^+$ with available lattice QCD data~\cite{Padmanath:2013zfa}. We report herein $S$-, $P$- and $D$-wave ground and radial-excited states for all channels mentioned. Since the total wave function of a $ccc$-baryon must be totally antisymmetric in order to fulfill the Pauli principle, there is no $S$-wave bound-state with total spin and parity $J^P=\frac{1}{2}^+$. Looking at Table~\ref{tab:Eccc}, we are predicting a mass of $4.80\,\text{GeV}$ for the ground state of the spectrum, which has quantum numbers $nL\,(J^P)=1S\,(\frac{3}{2}^+)$. The mass of the same state is predicted by lattice-QCD to be $4.76\,\text{GeV}$, which compares reasonably well with ours. A similar level of agreement between lattice and our calculation is repeated for each ground state of the reported $J^P$-channel in Table~\ref{tab:Eccc}.

Lattice QCD~\cite{Padmanath:2013zfa} reports two almost degenerate states in each channel with quantum numbers $J^P=\frac{1}{2}^+$, $\frac{3}{2}^+$ and $\frac{5}{2}^+$. They correspond to a different spin excitation because the three $J^P$ quantum numbers can be obtained when coupling a $D$-wave component with either $S=\frac{1}{2}$ or $\frac{3}{2}$. Table~\ref{tab:Eccc} shows the eigenstate of lowest mass, which corresponds to the coupling $L\otimes S=2\otimes3/2$. Our prediction for the remaining case, $L\otimes S=2\otimes1/2$, is $5407\,\text{MeV}$ which compares reasonably well with lattice results $5401\pm14\,\text{MeV}$, $5461\pm13\,\text{MeV}$ and $5460\pm15\,\text{MeV}$ for $\frac{1}{2}^+$, $\frac{3}{2}^+$ and $\frac{5}{2}^+$, respectively. Continuing with the postive-parity sector, the only radial excitation that can be compared with lattice is the $2S\,(\frac{3}{2}^+)$ state; our prediction, $5.29\,\text{GeV}$, is in fair agreement with the lattice one: $5.31\,\text{GeV}$. There is some mismatch, $\sim\!0.1\,\text{GeV}$, between our calculation and that using lattice-regularized QCD for negative-parity excited states (see Table~\ref{tab:Eccc} and Fig.~\ref{fig:Eccc}). 

Let us mention that the level of agreement (or disagreement) should be taken with some caution because our constituent quark model suffer of theoretical uncertainties that can be estimated to be $\pm50\,\text{MeV}$ when modifying $10\%$ the most sensitive model parameter. On the other hand, authors of Ref.~\cite{Padmanath:2013zfa} do not estimate all systematic uncertainties; that is to say, the lattice errors given in Table~\ref{tab:Eccc} and Fig.~\ref{fig:Eccc} are just statistical. With the lattice NRQCD action and parameters used in~\cite{Padmanath:2013zfa}, the systematic errors may be significant, especially for spin-dependent energy splittings. A calculation of the charmonium spectrum with the same lattice formulation can be found in Ref.~\cite{Liu:2012ze}, which could give an idea of the typical size of the systematic uncertainties in the lattice calculation.

Table~\ref{tab:Cccc} compares a significant sample of our prediction with the results reported by other theoretical formulations. For the ground state, $nL\,(J^P)=1S\,(\frac{3}{2}^+)$, our mass fairly agrees with the general trend, except for very few cases that predict a mass around $5.0\,\text{GeV}$. For the rest of the spectrum, the data reported by other approaches is quite sparse with big uncertainties in some cases, making difficult to perform a quantitative comparison. However, there are still some theoretical calculations~\cite{Vijande:2015faa, SilvestreBrac:1996bg, Migura:2006ep, Serafin:2018aih} where the reported spectrum is as complete as ours and one can deduce that the level of agreement is quite reasonable. 


\begin{table}[!t]
\caption{\label{tab:Eccb} Predicted masses, in MeV, of $\Omega_{ccb}$ baryons with total spin and parity $J^P=\frac{1}{2}^\pm$, $\frac{3}{2}^{\pm}$, $\frac{5}{2}^\pm$ and $\frac{7}{2}^+$. We compare our results with those obtained by lattice QCD in Ref.~\cite{Mathur:2018epb}.}
\begin{ruledtabular}
\begin{tabular}{cccc}
$J^P$ & $nL$ & This work & Ref.~\cite{Mathur:2018epb} \\
\hline
\tstrut
$\frac{1}{2}^+$ & $1S$ & $8004$ & $8005\pm13$ \\
                & $2S$ & $8455$ & - \\
$\frac{1}{2}^+$ & $1D$ & $8536$ & - \\
                & $2D$ & $8838$ & - \\
$\frac{3}{2}^+$ & $1S$ & $8023$ & $8026\pm13$ \\
                & $2S$ & $8468$ & - \\
$\frac{3}{2}^+$ & $1D$ & $8536$ & - \\
                & $2D$ & $8838$ & - \\
$\frac{5}{2}^+$ & $1D$ & $8536$ & - \\
                & $2D$ & $8838$ & - \\
$\frac{7}{2}^+$ & $1D$ & $8538$ & - \\
                & $2D$ & $8839$ & - \\[0.60ex]
\hline
\tstrut
$\frac{1}{2}^-$ & $1P$ & $8306$ & - \\
                & $2P$ & $8663$ & - \\
$\frac{3}{2}^-$ & $1P$ & $8306$ & - \\
                & $2P$ & $8663$ & - \\
$\frac{5}{2}^-$ & $1P$ & $8311$ & - \\
                & $2P$ & $8667$ & - \\
\end{tabular}
\end{ruledtabular}
\end{table}

\begin{figure*}[ht]
\centering
\includegraphics[clip,trim={0.0cm 0.0cm 0.0cm 0.0cm},width=0.8\textwidth, height=0.43\textheight]{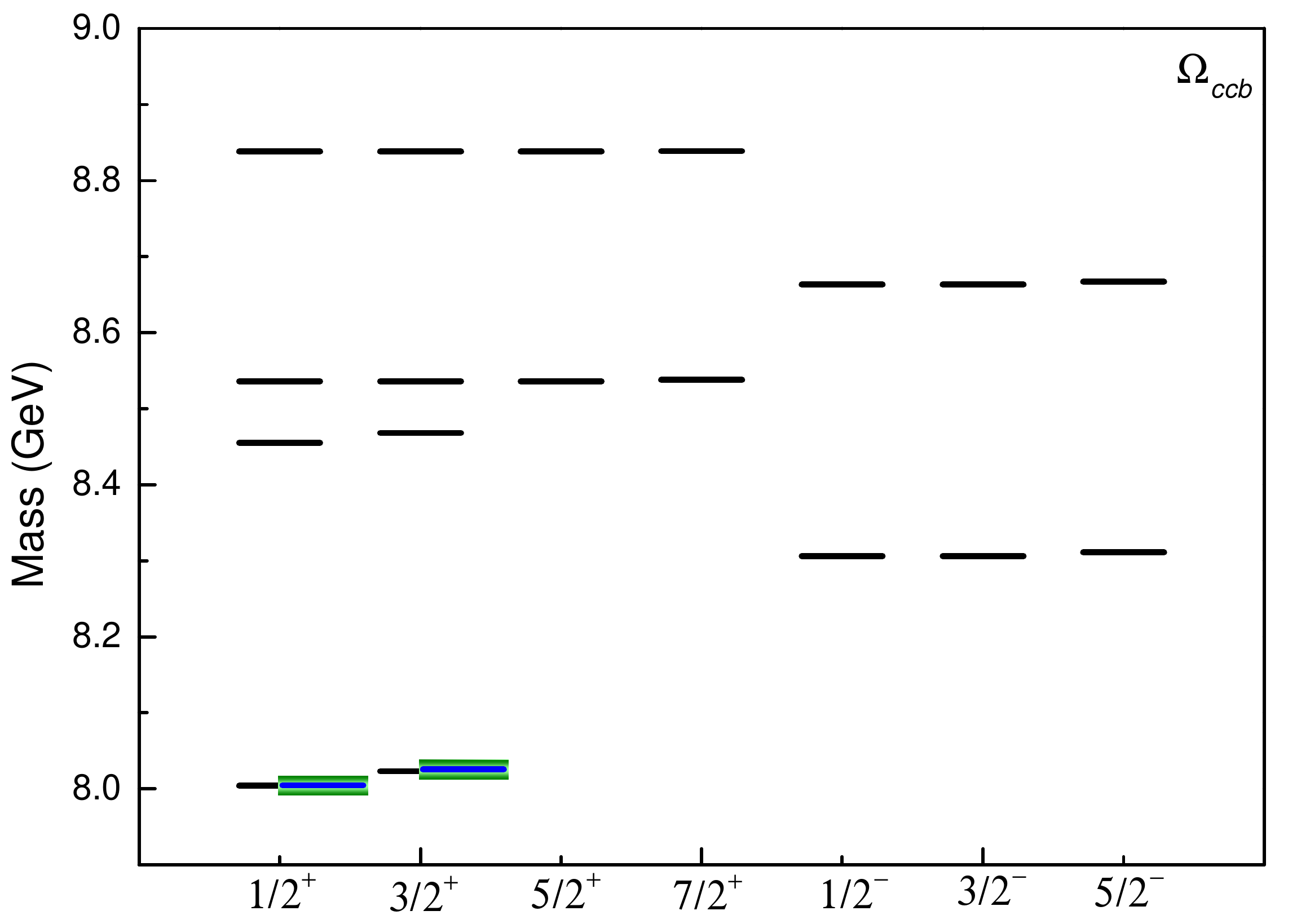}
\caption{\label{fig:Eccb} Spectrum of $\Omega_{ccb}$ baryons with total spin and parity $J^P=\frac{1}{2}^\pm$, $\frac{3}{2}^{\pm}$, $\frac{5}{2}^\pm$ and $\frac{7}{2}^+$. We compare our results (black solid lines) with those obtained by lattice QCD in Ref.~\cite{Mathur:2018epb} (blue lines with green boxes).}
\end{figure*}

\begin{table*}[!t]
\caption{\label{tab:Cccb} Predicted masses, in MeV, of $\Omega_{ccb}$ baryons by other theoretical approaches compared with our results.}
\begin{ruledtabular}
\begin{tabular}{ccccccccc}
$nL\, (J^P)$ & $1S\, (\frac{1}{2}^+)$ & $1S\, (\frac{3}{2}^+)$ & $2S\, (\frac{1}{2}^+)$ & $2S\, (\frac{3}{2}^+)$ & $1D\, (\frac{1}{2}^+)$  & $1P\, (\frac{1}{2}^-)$ & $1P\, (\frac{3}{2}^-)$ & $1P\, (\frac{5}{2}^-)$ \\
\hline
This work                   & $8004$ & $8023$ & $8455$ & $8468$ &$8536$ & $8306$ & $8306$ &$8311$ \\ 
\cite{Bhavin:2009bp}        & $8172\pm90$ & $8182\pm90$ & - & - & - & - & - & - \\
\cite{Martynenko:2007je}    & $8018$ & $8025$ & - & - & - & - & - & - \\
\cite{SilvestreBrac:1996bg} & $8019$ & $8056$ & $8450$ & $8465$ & $8528$ & $8316$ & $8316$ & $8331$ \\
\cite{Serafin:2018aih}      & $8301$ & $8301$ & $8600$ & $8600$ & $8647$ & $8491$ & $8491$ & $8491$ \\
\cite{Zhang:2009re}         & $7410\pm130$ & $7450\pm160$ & - & - & - & - & - & - \\
\cite{Wang:2011ae}          & $8230\pm130$ & $8230\pm130$ & - & - & - & $8360\pm130$ & $8360\pm130$ & - \\
\cite{Aliev:2012tt}         & $8500\pm120$ & - & - & - & - & - & - & - \\
\cite{Aliev:2014lxa}        & - & $8070\pm100$ & - & - & - & - & $8350\pm100$ & - \\
\cite{Jia:2006gw}           & $7980\pm70$ & - & - & - & - & - & - & - \\
\cite{LlanesEstrada:2011kc} & $8560\pm610$ & - & - & - & - & - & - & - \\
\cite{Qin:2019hgk}          & $7867$ & $7963$ & $8337$ & $8427$ & - & $8164$ & $8275$ & - \\
\cite{Yin:2019bxe}          & $8190$ & $8190$ & - & - & - & - & - & - \\
\cite{Zachary:2014zsb}      & $8007\pm9\pm20$ & $8037\pm9\pm20$ & - & - & - & - & - & - \\
\end{tabular}
\end{ruledtabular}
\end{table*}

{\bf The $\mathbf{\Omega_{ccb}}$ baryon sector.} Table~\ref{tab:Eccb} and Fig.~\ref{fig:Eccb} show our spectrum of $\Omega_{ccb}$ baryons. We are predicting two almost degenerate states with quantum numbers $nL\,(J^P)=1S\,(\frac{1}{2}^+)$ and $1S\,(\frac{3}{2}^+)$ and masses of the order of $8.0\,\text{GeV}$. These two constitute the ground of positive-parity $\Omega_{ccb}$ baryons and the agreement with recent lattice-QCD prediction~\cite{Mathur:2018epb} is remarkable. In this case, the lattice computation is based on (i) three different lattice spacings allowing precise results at the continuum limit; (ii) a relativistic formulation for the light, strange and charm quarks; (iii) a lattice NRQCD action for bottom quarks with  non-perturbatively tuned coefficients up to ${\cal O}(\alpha_s\,v^4)$; and (iv) a control of the statistical errors below percent level.

Let us now turn our attention to Table~\ref{tab:Cccb} in which we compare our results with those reported by other approaches. As one can see, there is no agreement about what would be the masses of the $1S\,(\frac{1}{2}^+)$ and $1S\,(\frac{3}{2})^+$ states. The mass splitting between them seems to be small, of the order of tens of MeV, but their absolute masses cluster around two different mean values, one at $8.0\,\text{GeV}$ and the other at around $8.2-8.3\,\text{GeV}$. This is quite puzzling, we do not have a sensible answer to this issue and then we suggest to continue investigating this sector. It is fair to note that there are some theoretical computations, mostly QCD sum rule predictions, where the reported masses of the $1S\,(\frac{1}{2}^+)$ and $1S\,(\frac{3}{2})^+$ states are quite different among each other and report large error bands. 

There are few computations~\cite{SilvestreBrac:1996bg, Serafin:2018aih, Qin:2019hgk} that provide a spectrum as complete as ours. The results presented in Ref.~\cite{SilvestreBrac:1996bg} are in reasonable agreement with our calculated masses, as shown in Table~\ref{tab:Cccb}. This could be because the formalism and quark-quark interactions are very similar despite the numerical tool and model parameters are different. The relativistic effects may have been (somehow) implemented in Refs~\cite{Serafin:2018aih, Qin:2019hgk}. The predicted states reported in ~\cite{Serafin:2018aih} are $0.2-0.3\,\text{GeV}$ higher than ours, and those reported by lattice-QCD~\cite{Mathur:2018epb}. Authors of Ref.~\cite{Qin:2019hgk} predict an spectrum in agreement with ours if we lift all their values by about $0.1\,\text{GeV}$. This indicates that, at least, mass splittings could be considered while disentangling experimentally the $\Omega_{ccb}$ spectrum.


\begin{table}[!t]
\caption{\label{tab:Ecbb} Predicted masses, in MeV, of $\Omega_{cbb}$ baryons with total spin and parity $J^P=\frac{1}{2}^\pm$, $\frac{3}{2}^{\pm}$, $\frac{5}{2}^\pm$ and $\frac{7}{2}^+$. We compare our results with those obtained by lattice QCD in Ref.~\cite{Mathur:2018epb}.}
\begin{ruledtabular}
\begin{tabular}{cccc}
$J^P$ & $nL$ & This work & Ref.~\cite{Mathur:2018epb} \\
\hline
\tstrut
$\frac{1}{2}^+$ & $1S$ & $11200$ & $11194\pm13$ \\
                & $2S$ & $11607$ & - \\
$\frac{1}{2}^+$ & $1D$ & $11677$ & - \\
                & $2D$ & $11955$ & - \\
$\frac{3}{2}^+$ & $1S$ & $11221$ & $11211\pm13$ \\
                & $2S$ & $11622$ & - \\
$\frac{3}{2}^+$ & $1D$ & $11677$ & - \\
                & $2D$ & $11955$ & - \\
$\frac{5}{2}^+$ & $1D$ & $11677$ & - \\
                & $2D$ & $11955$ & - \\
$\frac{7}{2}^+$ & $1D$ & $11688$ & - \\
                & $2D$ & $11963$ & - \\[0.60ex]
\hline
\tstrut
$\frac{1}{2}^-$ & $1P$ & $11482$ & - \\
                & $2P$ & $11802$ & - \\
$\frac{3}{2}^-$ & $1P$ & $11482$ & - \\
                & $2P$ & $11802$ & - \\
$\frac{5}{2}^-$ & $1P$ & $11569$ & - \\
                & $2P$ & $11888$ & - \\
\end{tabular}
\end{ruledtabular}
\end{table}

\begin{figure*}[ht]
\centering
\includegraphics[clip,trim={0.0cm 0.0cm 0.0cm 0.0cm},width=0.8\textwidth, height=0.43\textheight]{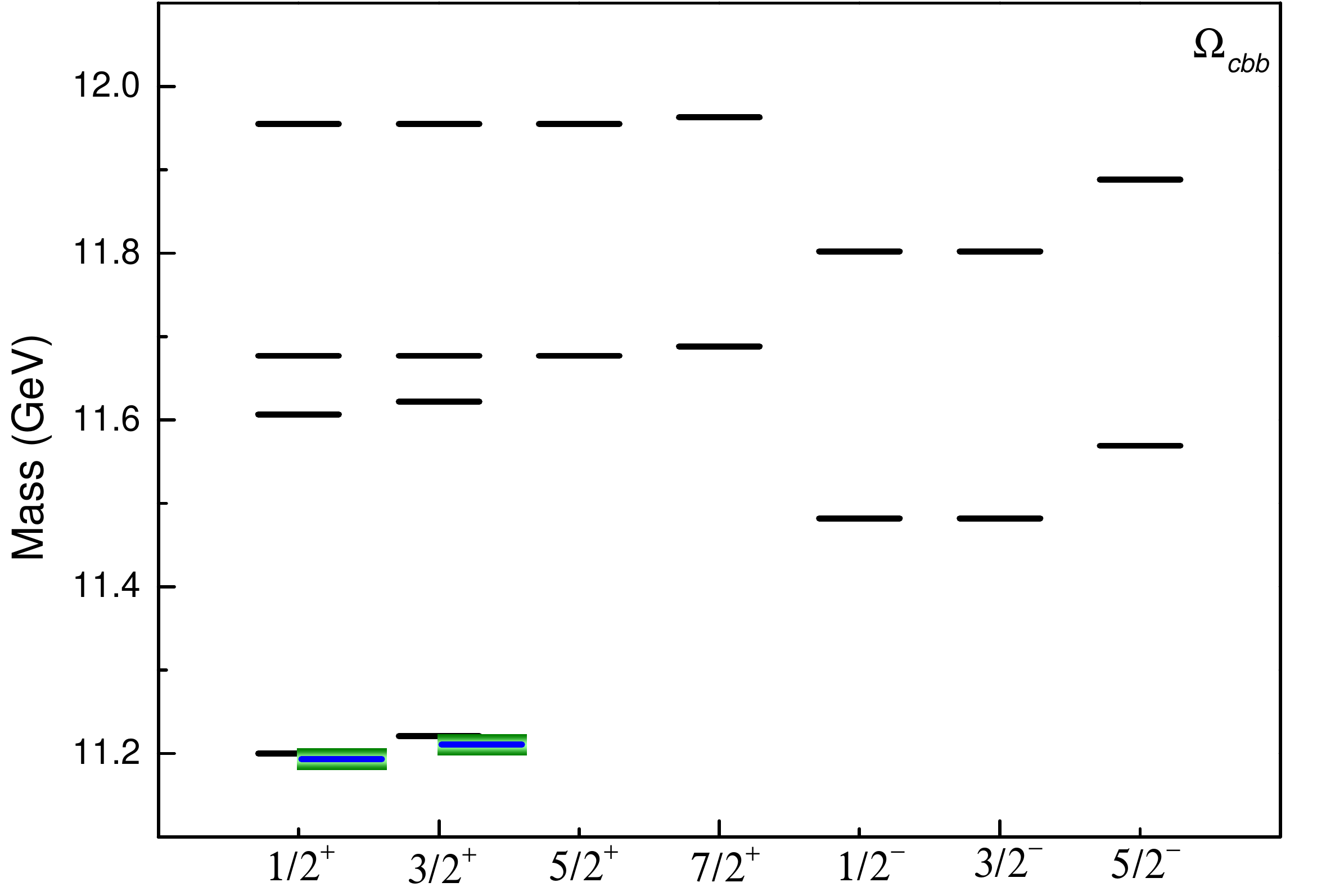}
\caption{\label{fig:Ecbb} Spectrum of $\Omega_{cbb}$ baryons with total spin and parity $J^P=\frac{1}{2}^\pm$, $\frac{3}{2}^{\pm}$, $\frac{5}{2}^\pm$ and $\frac{7}{2}^+$. We compare our results (black solid lines) with those obtained by lattice QCD in Ref.~\cite{Mathur:2018epb} (blue lines with green boxes).}
\end{figure*}

\begin{table*}[!t]
\caption{\label{tab:Ccbb} Predicted masses, in MeV, of $\Omega_{cbb}$ baryons by other theoretical approaches compared with our results.}
\begin{ruledtabular}
\begin{tabular}{ccccccccc}
$nL\, (J^P)$ & $1S\, (\frac{1}{2}^+)$ & $1S\, (\frac{3}{2}^+)$ & $2S\, (\frac{1}{2}^+)$ & $2S\, (\frac{3}{2}^+)$ & $1D\, (\frac{1}{2}^+)$ & $1P\, (\frac{1}{2}^-)$ & $1P\, (\frac{3}{2}^-)$ & $1P\, (\frac{5}{2}^-)$ \\
\hline
This work                   & $11200$ & $11221$ & $11607$ & $11622$ &$11677$ & $11482$ & $11482$ & $11569$ \\
\cite{Bhavin:2009bp}        & $11446\pm99$ & $11487\pm101$ & - & - & - & - & - & - \\
\cite{Martynenko:2007je}    & $11280$ & $11287$ & - & - & - & - & - & - \\ 
\cite{SilvestreBrac:1996bg} & $11217$ & $11251$ & $11625$ & $11643$ & $11718$ & $11524$ & $11524$ & $11598$ \\
\cite{Serafin:2018aih}      & $11218$ & $11218$ & $11585$ & $11585$ & $11626$ & $11438$ & $11438$ & $11601$ \\
\cite{Zhang:2009re}         & $10300\pm100$ & $10450\pm110$ & - & - & - & - & - & - \\
\cite{Wang:2011ae}          & $11500\pm110$ & $11490\pm110$ & - & - & - & $11620\pm110$ & $11620\pm110$ & - \\
\cite{Aliev:2012tt}         & $11730\pm160$ & - & - & - & - & - & - & - \\
\cite{Aliev:2014lxa}        & - & $11350\pm150$ & - & - & - & - & $11500\pm200$ & - \\ 
\cite{Jia:2006gw}           & $11190\pm80$ & - & - & - & - & - & - & - \\
\cite{LlanesEstrada:2011kc} & $11675\pm785$ & - & - & - & - & - & - & - \\
\cite{Qin:2019hgk}          & $11077$ & $11167$ & $11603$ & $11703$ & - & $11413$ & $11523$ & - \\
\cite{Yin:2019bxe}          & $11370$ & $11380$ & - & - & - & - & - & - \\
\cite{Zachary:2014zsb}      & $11195\pm8\pm20$ & $11229\pm8\pm20$ & - & - & - & 
- & - & - \\
\end{tabular}
\end{ruledtabular}
\end{table*}

{\bf The $\mathbf{\Omega_{cbb}}$ baryon sector.} Table~\ref{tab:Ecbb} and Fig.~\ref{fig:Ecbb} show our spectrum of $\Omega_{cbb}$ baryons and compare it with recent lattice-QCD predictions~\cite{Mathur:2018epb}. Our predicted masses for the $nL\,(J^P)=1S\,(\frac{1}{2}^+)$ and $1S\,(\frac{3}{2}^+)$ states are $11200\,\text{MeV}$ and $11221\,\text{MeV}$, respectively. They agree again with lattice QCD results which, for the reasons specified above, are considered quite robust and precise.

Table~\ref{tab:Ccbb} shows a sample of the $\Omega_{cbb}$ baryon spectrum computed by other formulations. We find in this sector a similar situation than the one already discussed for $\Omega_{ccb}$. There is not a clear consensus between the different approaches about what would be the masses of the $nL\,(J^P)=1S\,(\frac{1}{2}^+)$ and $1S\,(\frac{3}{2}^+)$ states. Looks like they converge to a value of around $11.2\,\text{GeV}$, but again quite different results with large uncertainties are given by QCD sum rules introducing some noise difficult to disentangle. As expected, the calculation of Ref.~\cite{SilvestreBrac:1996bg} is in fair agreement with ours. The comparison of our results with those reported by relativistic formulations reveals an opposite situation with respect the $\Omega_{ccb}$ baryon sector: the computation of Ref.~\cite{Serafin:2018aih} seems to agree with our results but, this time, we lost track with the masses reported in Ref.~\cite{Qin:2019hgk}.

Maybe not connected with the above, it is interesting to mention that the reported mass of Ref.~\cite{Jia:2006gw}, which is a model independent prediction based on the application of a non-relativistic effective field theory, agrees nicely with our result. Note too that the same level of agreement with the predictions of Ref.~\cite{Jia:2006gw} exists in all $\Omega_{QQQ}$ sectors, with $Q$ either a $c$- or $b$-quark. 


\begin{table}[!t]
\caption{\label{tab:Ebbb} Predicted masses, in MeV, of $\Omega_{bbb}$ baryons with total spin and parity $J^P=\frac{1}{2}^\pm$, $\frac{3}{2}^{\pm}$, $\frac{5}{2}^\pm$ and $\frac{7}{2}^+$. We compare our results with those obtained by lattice QCD in Ref.~\cite{Meinel:2012qz}.}
\begin{ruledtabular}
\begin{tabular}{cccc}
$J^P$ & $nL$ & This work & Ref.~\cite{Meinel:2012qz} \\
\hline
\tstrut
$\frac{1}{2}^+$ & $1D$ & $14894$ & $14938\pm18$ \\
                & $2D$ & $15175$ & - \\
$\frac{3}{2}^+$ & $1S$ & $14396$ & $14371\pm12$ \\
                & $2S$ & $14805$ & $14840\pm14$ \\
$\frac{3}{2}^+$ & $1D$ & $14894$ & $14958\pm18$ \\
                & $2D$ & $15175$ & - \\
$\frac{5}{2}^+$ & $1D$ & $14894$ & $14964\pm18$ \\
                & $2D$ & $15175$ & - \\
$\frac{7}{2}^+$ & $1D$ & $14894$ & $14969\pm17$ \\
                & $2D$ & $15175$ & - \\[0.60ex]
\hline
\tstrut
$\frac{1}{2}^-$ & $1P$ & $14688$ & $14706.3\pm9$ \\
                & $2P$ & $15016$ & - \\
$\frac{3}{2}^-$ & $1P$ & $14688$ & $14714\pm9$ \\
                & $2P$ & $15016$ & - \\
$\frac{5}{2}^-$ & $1P$ & $15038$ & - \\
                & $2P$ & $15284$ & - \\
\end{tabular}
\end{ruledtabular}
\end{table}

\begin{figure*}[ht]
\centering
\includegraphics[clip,trim={0.0cm 0.0cm 0.0cm 0.0cm},width=0.8\textwidth, height=0.43\textheight]{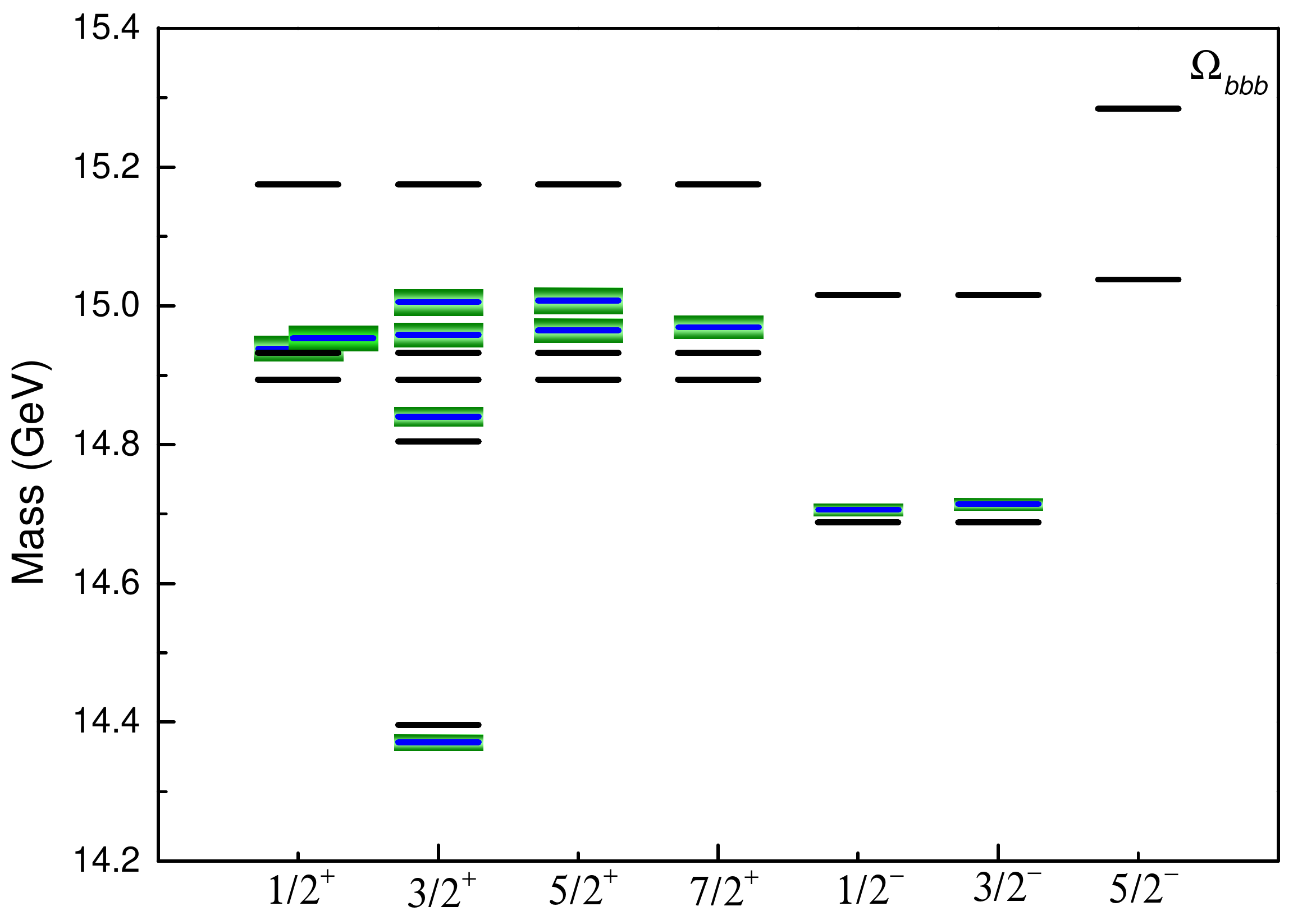}
\caption{\label{fig:Ebbb} Spectrum of $\Omega_{bbb}$ baryons with total spin and parity $J^P=\frac{1}{2}^\pm$, $\frac{3}{2}^{\pm}$, $\frac{5}{2}^\pm$ and $\frac{7}{2}^+$. We compare our results (black solid lines) with those obtained by lattice QCD in Ref.~\cite{Meinel:2012qz} (blue lines with green boxes).}
\end{figure*}

\begin{table*}[!t]
\caption{\label{tab:Cbbb} Predicted masses, in MeV, of $\Omega_{bbb}$ baryons by other theoretical approaches compared with our results.}
\begin{ruledtabular}
\begin{tabular}{ccccccc}
$nL\, (J^P)$ & $1S\, (\frac{3}{2}^+)$ & $2S\, (\frac{3}{2}^+)$ & $1D\, (\frac{1}{2}^+)$ & $1P\, (\frac{1}{2}^-)$ & $1P\, (\frac{3}{2}^-)$ & $1P\, (\frac{5}{2}^-)$ \\
\hline
This work                   & $14396$ & $14805$ & $14894$ & $14688$ & $14688$ & $15038$ \\
\cite{Bhavin:2009bp}        & $14566\pm122$ & - & - & - & - & - \\
\cite{Vijande:2015faa}      & $14371$ & $14848$ & $14954$ & $14713$ & $14713$ & $15125$ \\
\cite{Martynenko:2007je}    & $14569$ & - & - & - & - & - \\ 
\cite{SilvestreBrac:1996bg} & $14398$ & $14835$ & $14944$ & $14738$ & $14738$ & $15078$ \\
\cite{Serafin:2018aih}      & $14347$ & $14832$ & $14896$ & $14645$ & $14645$ & - \\
\cite{Zhang:2009re}         & $13280\pm100$ & - & - & - & - & - \\
\cite{Wang:2011ae}          & $14830\pm100$ & - & - & - & $14950\pm110$ & - \\
\cite{Aliev:2014lxa}        & $14300\pm200$ & - & - & - & $14900\pm200$ & - \\ 
\cite{Jia:2006gw}           & $14370\pm80$ & - & - & - & - & - \\
\cite{LlanesEstrada:2011kc} & $14935\pm735$ & - & - & - & - & - \\
\cite{Qin:2019hgk}          & $14370$ & $14980$ & - & - & $14771$ & - \\
\cite{Yin:2019bxe}          & $14570$ & - & - & - & - & - \\
\cite{Meinel:2010pw}        & $14371$ & - & - & - & - & - \\
\cite{Zachary:2014zsb}      & $14366\pm9\pm20$ & - & - & - & - & - \\
\end{tabular}
\end{ruledtabular}
\end{table*}

{\bf The $\mathbf{\Omega_{bbb}}$ baryon sector.} Table~\ref{tab:Ebbb} and Fig.~\ref{fig:Ebbb} compare our predicted masses of $\Omega_{bbb}$ baryons with available lattice QCD data~\cite{Meinel:2012qz}. Let us again highlight that such comparison should be taken with some care. On one hand, our quark model uncertainties could be large but, in principle, they should be smaller than in the other triply-heavy baryon sectors. On the other hand, one should keep in mind the particular features of the lattice QCD calculation which, this time, is characterized by extrapolations of the spectrum to realistic values of the light sea-quark masses and bottom quarks implemented through a lattice NRQCD action.

We are predicting a mass of $14.40\,\text{GeV}$ for the lowest state in the spectrum which has quantum numbers $nL(J^P)=1S(\frac{3}{2}^+)$. The mass of the same state is predicted by lattice to be $14.37\,\text{GeV}$. The agreement between lattice and our calculation for each ground state of the $J^P$-channels reported in Table~\ref{tab:Ebbb} is worse (but not dramatic) than in the $\Omega_{ccc}$ sector. The only radial excitation that can be compared with lattice is the $2S\,(\frac{3}{2}^+)$ state; our prediction, $14.81\,\text{GeV}$, is in fair agreement with the lattice one: $14.84\,\text{GeV}$. As in the $\Omega_{ccc}$ baryon sector, Table~\ref{tab:Ebbb} shows the eigenstate of lowest mass with quantum numbers $J^{P}=\frac{1}{2}^+$, $\frac{3}{2}^+$ and $\frac{5}{2}^+$, and corresponding to the coupling $L\otimes S=2\otimes3/2$. Our prediction for the remaining case, $L\otimes S=2\otimes1/2$, is $14932\,\text{MeV}$ which compares reasonably well with lattice results $14953\pm18\,\text{MeV}$, $15005\pm19\,\text{MeV}$ and $15007\pm19\,\text{MeV}$ for $\frac{1}{2}^+$, $\frac{3}{2}^+$ and $\frac{5}{2}^+$, respectively.

We turn now our attention to Table~\ref{tab:Cbbb}, that compares a sample of the results predicted by our quark model with the ones reported by other theoretical approaches. There are again few results, mostly reported from QCD sum rules, which scatter the general trend of the ground state mass. If one keeps these results out of the average, one realizes that the $1S\,(\frac{3}{2}^+)$ state should have a mass around $14.40\,\text{GeV}$, which agrees with ours. It is nice to observe fair agreement between our reported results and the ones computed in Refs.~\cite{SilvestreBrac:1996bg, Serafin:2018aih}. A spectrum as complete as ours is also reported in Ref.~\cite{Vijande:2015faa}, one can see that a global agreement with our prediction is found. However, Ref.~\cite{Vijande:2015faa} uses a linear confining interaction between quarks which produces, in general, larger masses for higher radial and orbital excitations.

Let us finish this section remarking that although the production of triply-bottom baryons as well as their identification could be extremely difficult, we consider that the experimental hunt of these states must be pursued harder. The $\Omega_{bbb}$ system is theoretically the most interesting one of all studied herein because its triply bottom-quark content makes it the most non-relativistic, conventional few-body bound-state system that QCD can form nowadays.


\section{Summary}
\label{sec:summary}

The study of heavy quarkonia has revealed very useful to examine some relevant QCD's properties and fundamental parameters, without taking into account the usual light-quark complications. Therefore, the triply-heavy baryons may provide a complementary window for the understanding of QCD. Moreover, as in charmonium and bottomonium, there is no restriction of finding exotic candidates in the spectra of $QQQ$-baryon and thus a reliable prediction of conventional triply-heavy baryons is interesting in order to deliver a template from which to compare the future experimental findings.

Within a constituent quark model approach, we compute the ground- and excited-state energies of triply-heavy baryons with quantum numbers $J^P=\frac{1}{2}^{\pm}$, $\frac{3}{2}^{\pm}$, $\frac{5}{2}^{\pm}$ and $\frac{7}{2}^+$. No quark model parameter has been changed in this study, they were fitted during the last decade to reproduced successfully a diverse array of heavy quarkonium properties such as masses but also electromagnetic, strong and weak decays and reactions.

We solve the non-relativistic $3$-body bound state equation by means of a variational method in which the wave function solution is expanded using infinitesimally-shifted Gau\ss ians. This new set of basis functions makes the calculation of $3$-body matrix elements easier without the laborious Racah algebra. The Gau\ss ian ranges are taken in a geometric progression enabling its optimization employing a small number of free parameters. Moreover, the geometric progression is dense at short distances, so that it allows the description of the dynamics mediated by short range potentials. The fast damping of the Gau\ss ian tail is not a problem, since we can choose the maximal range much longer than the hadronic size.

There are no experimental data related with triply-heavy baryons. Our spectrum for $\Omega_{ccc}$ and $\Omega_{bbb}$ sectors can be compared with available lattice-regularised QCD computations. One can state that there is a reasonable agreement with lattice for the ground state of all studied $J^P$-channels; however, discrepancies are found for excited states. Some of them can be related with our limitations and theoretical uncertainties but lattice-regularized computations have their own issues such as the use of NRQCD actions for heavy quarks and the fact of not addressing all the systematic uncertainties. We have also compared our results with those computed by other theoretical approaches reaching the conclusion that there is a general trend about what would be the mass of the lowest state in the spectra of $\Omega_{ccc}$ and $\Omega_{bbb}$. The prediction of higher excited states have not been done in a very systematic way by many theoretical formulations; however, we have commented on those cases that present a spectrum as complete as ours.

The $\Omega_{ccb}$ and $\Omega_{cbb}$ sectors have been less explored by lattice QCD. Only masses for the $nL\,(J^P)=1S\,(\frac{1}{2}^+)$ and $1S\,(\frac{3}{2}^+)$ states were reported and they agree with our results. A wide array of theoretical predictions is available for the $\Omega_{ccb}$ and $\Omega_{cbb}$ sectors. If one could discard the results predicted by QCD sum rules, a general trend of agreement among very different approaches is revealed for the average mass of the $nL\,(J^P)=1S\,(\frac{1}{2}^+)$ and $nL\,(J^P)=1S\,(\frac{3}{2}^+)$ states. This value is compatible with our prediction. Again, few theoretical formulations report a complete spectrum of low-lying excited states; when available, we have compared them with our calculation.

It is interesting to remark herein that the spectra of $\Omega_{ccc}$, $\Omega_{ccb}$, $\Omega_{cbb}$ and $\Omega_{bbb}$ baryons are revealed to be quite reach in just an energy region of $1\,\text{GeV}$ above the corresponding ground state. Therefore, we encourage the design of experiments able to detect this kind of particles because the reward could be high and, as mentioned above, triply-heavy baryons are ideally suited to study QCD as it has been the case for heavy quarkonia.

Finally, after this computation, a possible worthwhile direction would be coupling the na\"ive triply-heavy baryons presented herein with their closer baryon-meson thresholds using the $^3P_0$ decay model as the mechanism connecting both $3$- and $5$-quark sectors. Mass-shifts, decay widths and all kind of scattering phenomena will then be available to study exotic structures. A similar procedure has been followed by some of us in the heavy quark meson sector with a great level of success.


\section{Acknowledgments}

The authors would like to thank L. He, D.R. Entem, F. Fern\'andez, and C.D. Roberts for their support and informative discussions. Work supported by: China Postdoctoral Science Foundation Grant no. 2019M650617; National Natural Science Foundation of China under Grant nos. 11535005 and 11775118; and by Spanish Ministerio de Econom\'ia, Industria y Competitividad under contracts no. FPA2017-86380-P and FPA2016-77177-C2-2-P. P.G.O. acknowledges the financial support from Spanish MINECO's Juan de la Cierva-Incorporaci\'on programme, Grant Agreement No. IJCI-2016-28525.


\bibliography{THB}

\end{document}